\documentstyle[preprint,aps]{revtex}
\tighten
\begin{document}
\draft
\title{Quantum Monte Carlo calculations of pion scattering from Li}
\author{T.-S. H. Lee and R.B. Wiringa}
\address{ Physics Division, Argonne National Laboratory,
         Argonne, Illinois 60439}
\date{\today}
\maketitle

\begin{abstract}
We show that the neutron and proton transition densities
predicted by recent Quantum Monte Carlo calculations
for $A=6,7$ nuclei are consistent with pion
scattering from $^6$Li and $^7$Li at energies
near the $\Delta$ resonance.
This has provided a microscopic understanding of the enhancement
factors for quadrupole excitations,
which were needed to describe pion inelastic scattering within
the nuclear shell model of Cohen and Kurath.
\end{abstract}

\pacs{PACS numbers: 21.60.Ka, 25.80.Ek}

Quantum Monte Carlo (QMC) methods have been successfully
developed to predict the properties of low-lying states of light nuclei
starting with realistic two- and three-nucleon potentials\cite{PPCPW97}.
While the reproduction of the energy spectra is 
the essential first step in such an effort, 
the dynamical content of the resulting
nuclear wave functions must be tested against various reactions.
This has been achieved so far mainly by considering electroweak
processes\cite{WS98,LWW99}. In this paper we report on an additional test
by using pion elastic and inelastic scattering.

Let us first briefly review the status of our understanding of
pion inelastic scattering at medium energies( 80 MeV $< E_{lab} < $ 
300 MeV). Because of the excitation of the $\Delta$
resonance, the pion-nucleus interactions in this energy region
are dominated by the strong absorption mechanism. 
Consequently, the pion-nucleus inelastic
scattering leading to discrete final nuclear states
can be described by the Distorted Wave Impulse 
Approximation (DWIA).  This has been well established
\cite{lee71,edw71,lee74,leekurath,olm80,see81,see82,mor87,oak87,rit91} 
in very extensive investigations of the data from meson factories.
Following the momentum-space
approach\cite{leekurath}, the inelastic scattering amplitude can be 
written as
\begin{equation}
T_{fi}(\vec{k}_0^{\, \prime},\vec{k}_0)
=\int d\vec{k}^{\,\prime} \int d\vec{k} 
\chi^{(-)*}_{\vec{k}_0,f}(\vec{k}^{\,\prime})
U_{fi}(\vec{k}^{\,\prime},\vec{k}) \chi^{(+)}_{\vec{k}_0,i}(\vec{k}) \ ,
\end{equation}
where the $\vec{k}$'s are pion-nucleus relative momenta in the
pion-nucleus center of mass frame, and 
the distorted waves $\chi^{(\pm)}$ are generated from
an optical potential which is adjusted to fit the pion-nucleus
elastic scattering. 

The nuclear excitations are contained in the transition potential
$U_{fi}$.  It can be calculated from the $\pi N$
scattering t-matrix and nuclear transition form factors.
The details are given in Ref.\cite{leekurath}. 
To simplify the presentation, 
the spin-isospin variables will be suppressed here.
Then the transition potential can be written as
\begin{eqnarray}
U_{fi}(\vec{k}^{\,\prime},\vec{k}) = 
t_{\pi n}(\vec{k}^{\,\prime}
,\vec{k},\omega_0) f^{-1/2}_{fi}(\vec{q})
+t_{\pi p}(\vec{k}^{\,\prime}
,\vec{k},\omega_0) f^{1/2}_{fi}(\vec{q})\, ,
\end{eqnarray}
where $\vec{q}=\vec{k}^{\,\prime}-\vec{k}$,
$t_{\pi n}(t_{\pi p})$ is an appropriately parameterized
pion-neutron (pion-proton) scattering amplitude, and $\omega_0$
is the collision energy calculated from using the 
fixed-scatterer approximation.
With $t_3=-1/2, 1/2$ denoting neutron and 
proton respectively, the nuclear transition form factors are
defined by
\begin{eqnarray}
f^{t_3}_{fi} (\vec{q})=
\int d\vec{r} e^{-i\vec{r}\cdot \vec{q}} 
\rho_{fi}^{t_3}(\vec{r})\, ,
\end{eqnarray}
with the transition densities defined by
\begin{eqnarray}
\rho_{fi}^{t_3}(\vec{r}) = < \Psi_f \mid 
\frac{1}{A}\sum_{i=1,A} 
\delta(\vec{r}-\vec{r}_i)\frac{1+ 2t_3 \tau_z(i)}{2}
\mid \Psi_i >\, , 
\end{eqnarray}
where $\tau_z $ is the z-component of the nucleon isospin operator, and
$\Psi_i$ and $\Psi_f$ are the initial and final nuclear
states respectively.
The transition dynamics can be better understood from the
multipole expansion of transition densities
\begin{eqnarray}
\rho_{fi}^{t_3}(\vec{r}) = \sum_{KM} \frac{1}{r^2}Y_{KM}(\hat{r}) 
<J_f M_f \mid J_i K M_i M> F^{fiK}_{K0,t_3}(r)\, ,   
\end{eqnarray} 
Here we recall the notation of Ref.\cite{leekurath} to include
possible spin transitions $S=0,1$ in
$ F^{fiJ}_{KS,t_3}(r)$, where $K$ denotes the orbital angular momentum
transition and $\vec{J} =\vec{K} + \vec{S}$ is the total angular momentum
transfer. 
   
In the DWIA study\cite{leekurath} for 1p-shell nuclei, 
the transition densities were 
calculated from the shell model of Cohen and Kurath\cite{cohen}:
\begin{eqnarray}
F^{fiJ}_{KS,t_3}(r) &=& \sum_{\alpha\beta}A_{J(KS)t_3}(\alpha,\beta;fi) 
(4\pi j_\alpha)^{1/2} 
<(l_\alpha 1/2)j_\alpha \mid\mid [Y_K(\hat{r})\times \sigma_S]_J
\mid\mid(l_\beta 1/2)j_\beta > \nonumber \\
&\times& R_{n_\alpha l_\alpha}(r) R_{n_\beta, l_\beta}(r)
\end{eqnarray}
with $\sigma_0=1$, $\sigma_1 =\vec{\sigma}$, and
\begin{eqnarray}
A_{J(KS)t_3}(\alpha,\beta;fi) 
= < \Psi_f\mid\mid [b^\dagger_{\alpha t_3} \times 
h^\dagger_{\beta t_3} ]_{J[KS]}  \mid\mid \Psi_i > \ ,
\end{eqnarray}
where $\alpha(n_\alpha, l_\alpha, j_\alpha)$ denotes the 
single particle orbitals, $R_\alpha(r)$
is the radial wave function, and $b^\dagger_\alpha$ and 
$h^\dagger_\beta$ are the creation operators for the 
particle and hole
states respectively.

It was found that the pion inelastic scattering from 1p-shell
nuclei can be described by using the above shell-model input only
when the quadrupole excitation component $J(KS)=2(20)$ is enhanced by
a factor $E_N \sim 2$. In Ref.\cite{leekurath},
these enhancement factors were estimated from a systematic analysis of
$B(E2)$ transitions from 1p-shell nuclei. For proton
excitation, this enhancement factor is consistent with what is needed
for explaining $B(E2)$ values. For $Z=N$ nuclei, 
one can assume
that the neutron excitation also has the same enhancement because
of isospin invariance.
However, for the
$N\neq Z$ nuclei, such as $^7$Li, 
the enhancement factors for neutron excitations
can not be obtained without making some additional assumptions.
The predicted pion inelastic cross sections thus are not well
justified theoretically. Furthermore, it would be desirable if
the calculations do not include any enhancement factors. This
however is very difficult, if not impossible, in practice 
within the shell model since the collective
quadrupole excitations can only be described by a very large
model space.

In this work, we calculate the transition densities for a given
multipolarity by using wave functions from the recent QMC
calculations for light nuclei.
The input, Eq.(6), to the DWIA calculations for pion inelastic 
scattering is then defined by the following matrix element
\begin{eqnarray}
F^{fiJ}_{KS,t_3}(r) =\frac{ 
<\Psi_{J_f M_f} \mid \sum_{i=1,A}\delta(r-r_i)r^K_i
[Y_K(\hat{r}_i)\times \sigma_S]_{JM} \frac{1+2t_3\tau(i)}{2}
\mid \Psi_{J_i M_i} >}{<J_f M_f \mid J_i J M_i M >}
\end{eqnarray}
The QMC calculations use a realistic Hamiltonian containing the Argonne 
$v_{18}$ two-nucleon~\cite{WSS95} and Urbana IX~\cite{PPCW95}
three-nucleon potentials, which we refer to as the AV18/UIX model.
Both variational (VMC) and Green's function (GFMC) Monte Carlo
calculations have been made for light nuclei~\cite{PPCPW97}.
The AV18/UIX model reproduces the experimental binding energies and 
charge radii of $^3$H, $^3$He, and $^4$He, in the numerically exact 
GFMC calculations, but underbinds $^6$Li and $^7$Li by 2--5\%.
The variational Monte Carlo (VMC) energies are 2\% above the GFMC results
for $A=3,4$ nuclei and 10\% above for $A=6,7$.
However, the known excitation spectra are well reproduced by both the
VMC and GFMC calculations, as are the charge radii.
The VMC and GFMC calculations also produce very similar one-body densities, 
while two-nucleon density distributions differ by less than 10\%.

The VMC wave functions have been used successfully to describe the 
elastic and transition electromagnetic form factors for $^6$Li~\cite{WS98}
without introducing effective charges.
They have also given an excellent absolute prediction for the spectroscopic 
factors in $^7$Li($e,e^\prime p$) reaction~\cite{LWW99}.
Consequently we expect the VMC wave functions to give a good estimate 
for both the elastic and transition densities required in pion scattering
calculations.

The variational wave function for $A=6,7$ nuclei used here is the trial
wave function, $\Psi_T$, that serves as the starting point for the GFMC
calculations.
It has the general form
\begin{equation}
     |\Psi_T\rangle = \left[ 1 + \sum_{i<j<k} \tilde{U}^{TNI}_{ijk} \right]
              \left[ {\cal S}\prod_{i<j}(1+U_{ij}) \right] |\Psi_J\rangle \ ,
\end{equation}
where $U_{ij}$ and $\tilde{U}^{TNI}_{ijk}$ are two- and three-body
correlation operators and the Jastrow wave function
$|\Psi_J\rangle$ is given by
\begin{eqnarray}
  |\Psi_J\rangle &=& {\cal A} \left\{
     \prod_{i<j<k \leq 4}f^c_{ijk}
     \prod_{i<j \leq 4}f_{ss}(r_{ij})
     \prod_{k \leq 4 < l \leq A} f_{sp}(r_{kl}) \right. \nonumber\\
  && \left.  \sum_{LS} \Big( \beta_{LS[n]} \prod_{4 < l < m \leq A} 
     f^{LS[n]}_{pp}(r_{lm})
     |\Phi_{A}(LS[n]JMTT_{3})_{1234:56\ldots A}\rangle \Big) \right\} \ .
\end{eqnarray}
The ${\cal S}$ and ${\cal A}$ are symmetrization and antisymmetrization
operators, respectively.
The central pair and triplet correlations $f_{xy}(r_{ij})$ and
$f^c_{ijk}$ are functions of relative positions only; the subscripts $xy$
denote whether the particles are in the s- or p-shell.
The $|\Phi_A(LS[n]JMTT_{3})\rangle$ is a single-particle wave function with
orbital angular momentum $L$, spin $S$, and spatial symmetry $[n]$ 
coupled to total angular
momentum $J$, projection $M$, isospin $T$, and charge state $T_{3}$:
\begin{eqnarray}
 &&  |\Phi_{A}(LS[n]JMTT_{3})_{1234:56\ldots A}\rangle =
     |\Phi_{\alpha}(0 0 0 0)_{1234} \prod_{4 < l\leq A}
     \phi^{LS}_{p}(R_{\alpha l}) \nonumber \\
 &&  \left\{ [ \prod_{4 < l\leq A} Y_{1m_l}(\Omega_{\alpha l}) ]_{LM_L[n]}
     \times [ \prod_{4 < l\leq A} \chi_{l}(\case{1}{2}m_s) ]_{SM_S}
     \right\}_{JM} 
     \times [ \prod_{4 < l\leq A} \nu_{l}(\case{1}{2}t_3) ]_{TT_3}\rangle \ .
\end{eqnarray}
Particles 1--4 are placed in an $\alpha$ core with only spin-isospin degrees of
freedom, denoted by $\Phi_{\alpha}(0000)$, while particles 5--A are placed in
$p$-wave orbitals $\phi^{LS}_{p}(R_{\alpha l})$ that are functions of the
distance between the center of mass of the $\alpha$ core and particle $l$.
Different amplitudes $\beta_{LS[n]}$ are mixed to obtain an optimal wave
function by means of a small-basis diagonalization.
For $^6$Li, the $(J^{\pi};T)=(1^+;0)$ ground state is predominantly a $^3$S[2]
amplitude, with small admixtures of $^3$D[3] and $^1$P[11] components, while 
the $(3^+;0)$ first excited state is pure $^3$D[3].
For $^7$Li, the $(J^{\pi};T)=(\case{3}{2}^-;\case{1}{2})$ ground and 
$(\case{1}{2}^-;\case{1}{2})$ first excited states are predominantly
$^2$P[3], with small admixtures of $^{2,4}$P[21], $^{2,4}$D[21],
and $^2$S[111] components.
The $(J^{\pi};T)=(\case{7}{2}^-;\case{1}{2})$ and 
$(\case{5}{2}^-;\case{1}{2})$ excited states are predominantly $^2$F[3],
again with small admixtures of $^{2,4}$P[21] and $^{2,4}$D[21] components.
Mixing parameter values are given in Ref.~\cite{PPCPW97}.

The two-body correlation operator $U_{ij}$ is defined as:
\begin{equation}
     U_{ij} = \sum_{p=2,6} \left[ \prod_{k\not=i,j}f^p_{ijk}({\bf r}_{ik}
              ,{\bf r}_{jk}) \right] u_p(r_{ij}) O^p_{ij} \ ,
\end{equation}
where the $O^{p=2,6}_{ij}$ = ${\bbox \tau}_i\cdot {\bbox \tau}_j$,
${\bbox \sigma}_i\cdot{\bbox \sigma}_j$,
${\bbox \sigma}_i\cdot{\bbox \sigma}_j {\bbox \tau}_i\cdot {\bbox \tau}_j$,
$S_{ij}$, and $S_{ij}{\bbox \tau}_i\cdot {\bbox \tau}_j$.
The six radial functions $f_{ss}(r)$ and $u_{p=2,6}(r)$ are obtained from
two-body Euler-Lagrange equations with variational parameters~\cite{W91}.
The $f_{sp}$ and $f^{LS[n]}_{pp}$ correlations are similar to $f_{ss}$ for
small separations, but include parameterized long-range tails.
The parameters used in constructing these two-body correlations, as well as
the description of the three-body correlation operator $\tilde{U}^{TNI}_{ijk}$
and the operator-independent three-body correlations $f^c_{ijk}$ and
$f^{p}_{ijk}$ are given in Ref.~\cite{PPCPW97}.

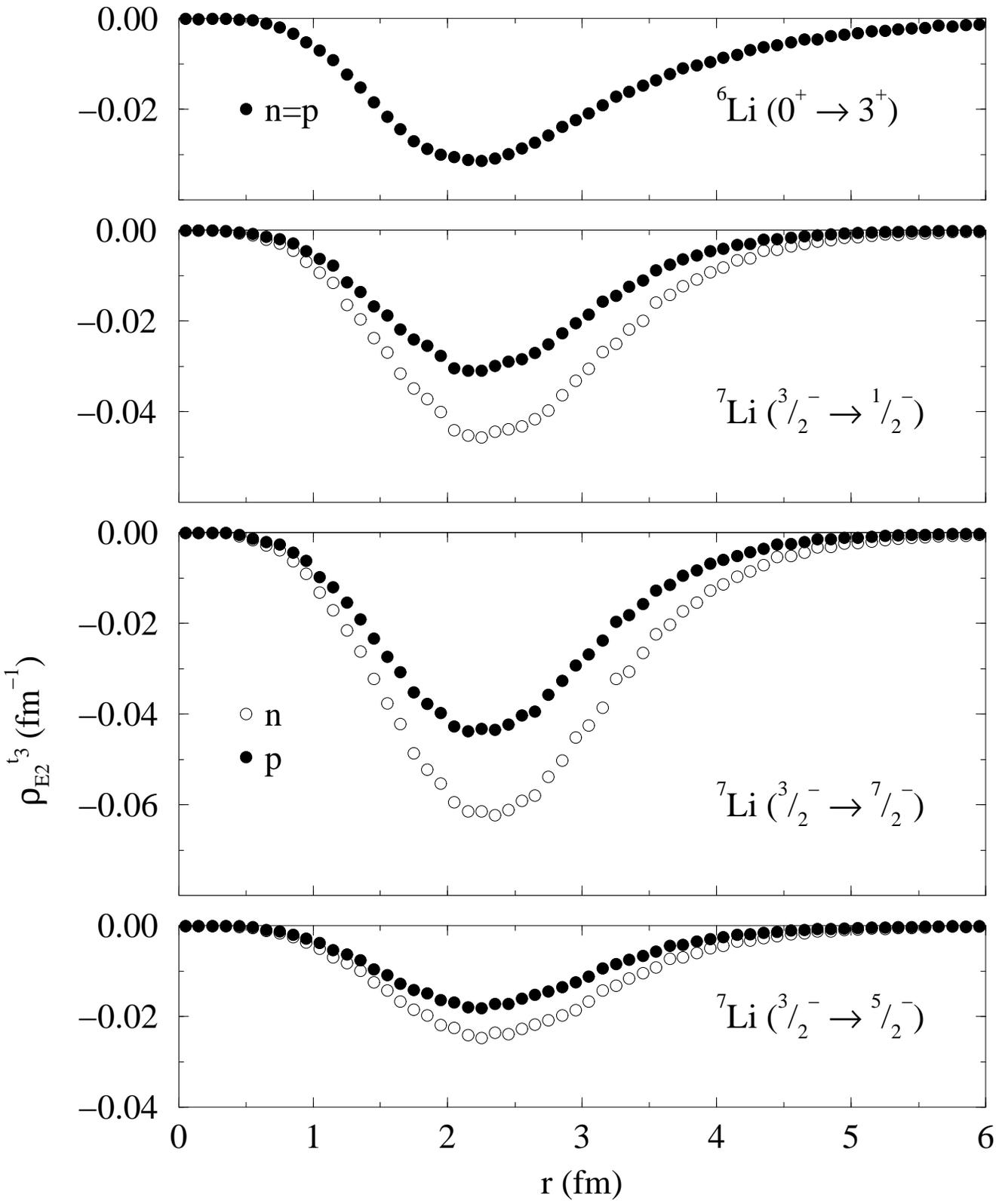
\begin{figure}
\caption{The transition densities, $\rho^{n,p}_{E2}$, for 
$^6$Li and $^7$Li.}
\label{fig:density}
\end{figure}

In Ref.~\cite{leekurath} it was found that inelastic transitions induced
by pion scattering are dominated by the quadrupole transition $J(KS)=2(20)$.
In our QMC calculations, we evaluate the quadrupole transition density:
\begin{equation}
\rho^{t_3}_{E2} (r) = \frac{ \sqrt{2 J_f +1}
 \langle \Psi_{J_f M_f} \mid \sum_{i=1,A} \delta(r-r_i) r^2_i Y^M_2(\hat{r}_i)
 \frac{1+2t_3\tau(i)}{2} \mid \Psi_{J_i M_i} \rangle }
 {\langle J_f M_f \mid J_i 2 M_i M \rangle} \ .
\end{equation}
These neutron and proton transition densities are shown in 
Fig.~\ref{fig:density} for four transitions in $^6$Li and $^7$Li. 
The integrated $B(E2\uparrow,t_3)$ values,
\begin{equation}
B(E2\uparrow,t_3) = 
    \frac{ \mid \int \rho^{t_3}_{E2} (r) d^3r \mid ^2}{2J_i + 1} \ ,
\end{equation}
are given in Table~\ref{tab:be2}, where they are compared to
the experimental proton values obtained from $(e,e^\prime)$
scattering and Coulomb excitation experiments~\cite{yen,vermeer,la66}.
Our evaluations are made with 160 000 Monte Carlo samples for transitions 
in $^6$Li and 120 000 Monte Carlo samples for transitions in $^7$Li.
This number of samples is sufficient to give statistical uncertainties that 
are as small or smaller than the errors on the experimental $B(E2)$ values.
Other $J(KS)$ transition amplitudes can contribute to pion inelastic
scattering, particularly in the $^6$Li case, and are evaluated in a similar 
manner.

We see from Fig.~\ref{fig:density} and Table~\ref{tab:be2} that 
the predicted differences between the neutron and proton excitations in
$^7$Li are very significant.
Such differences can be most effectively verified by using an 
important characteristic of pion scattering at energies near the $\Delta$
excitation. At a typical energy $E_\pi =164$ MeV, one finds that
the $\pi N$ amplitude in Eq.(2) has an interesting ratio
$\mid t_{\pi^+p}/t_{\pi^+n}\mid =\mid t_{\pi^-n}/t_{\pi^-p}\mid \sim 3$.
Consequently, the $\pi^+$ scattering is dominated by the proton excitations 
while $\pi^-$ scattering is dominated by neutron excitations. The agreement 
with both the $\pi^+$ and $\pi^-$ data will be a nontrivial test of the
QMC wave functions. Thus the present study is complementary to
that of Ref.~\cite{WS98} using electron scattering which mainly probes
the proton excitations. 

We first investigate pion scattering from $^6$Li. 
Here data for $E_\pi= 100$, 180 and 240 MeV~\cite{rit91} are available 
for testing the energy-dependence of our predictions.
The pion optical potential and the $\pi N$ t-matrix are taken from
Ref.\cite{leekurath}.
The $\pi N$ amplitudes we employ are taken from
the Karlsruhe-Helsinki analysis\cite{kh80},
and differ only slightly with the more recent VPI analysis\cite{vpi},
mainly in the $S_{11}$ partial wave (as discussed in Ref.\cite{svarc}).
This  partial wave and the other non-$P_{33}$ partial waves are much
weaker than the $P_{33}$ channel in the energy region of interest near the
$\Delta$ excitation.
For the present exploratory investigation, we therefore do not make any
effort to improve the optical potential employed in Ref.\cite{leekurath}.
Such an improvement is probably needed in the future when the data at
low energies and for the spin observales are investigated.

With transition densities calculated from QMC, there are no
adjustable parameters in our DWIA calculations.
In Table~\ref{tab:be2}, we see that the calculated $B(E2)$ for the
transition to the $(3^+, T=0)$ state is in excellent agreement with the data.
This is a significant improvement over the shell-model prediction which
required a large enhancement factor $E_p=E_n=2.5$ to reproduce the $B(E2)$
data, as discussed in Ref.\cite{leekurath}.
We thus expect a similar improvement in pion scattering calculations.

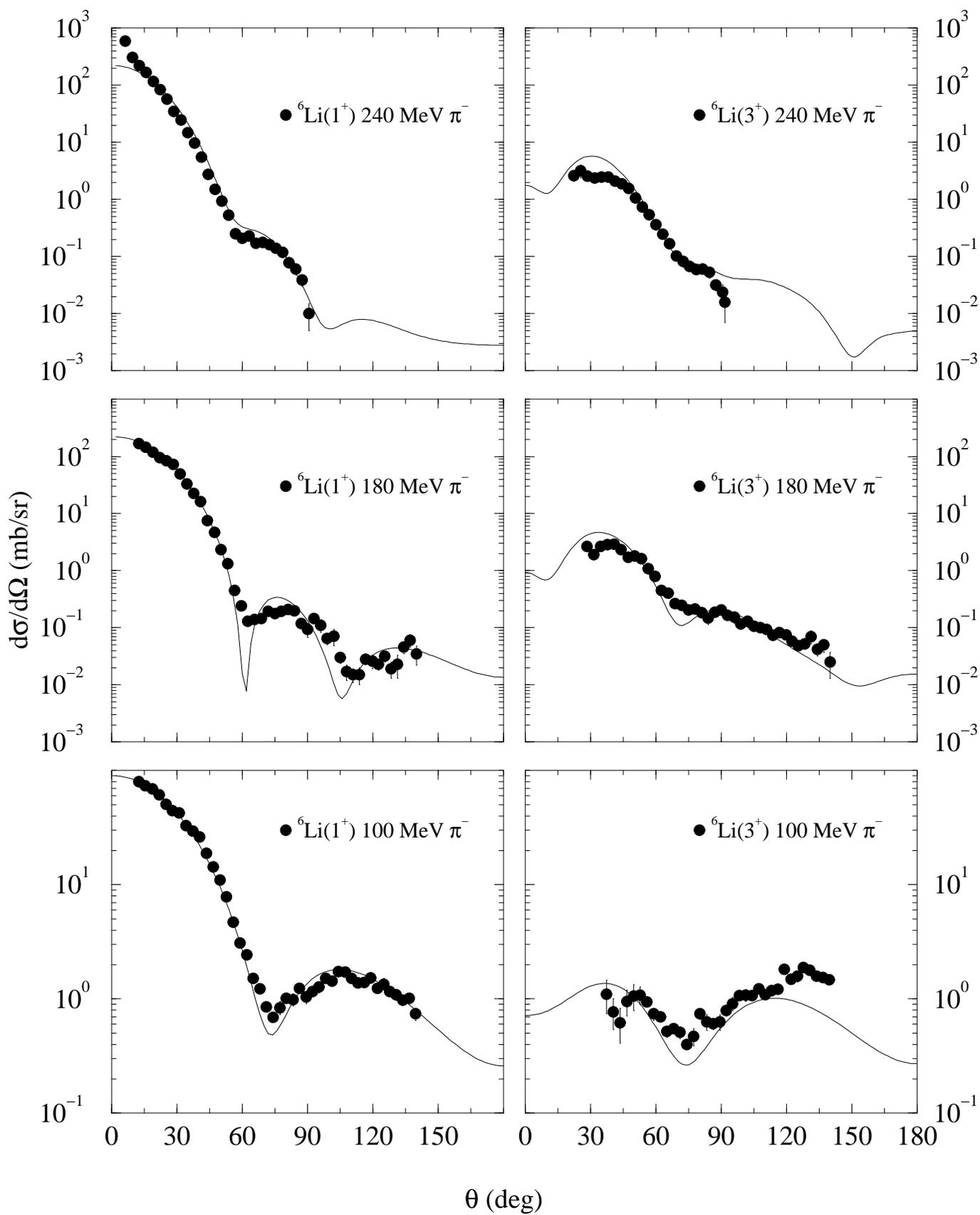
\begin{figure}
\caption{Differential cross sections for $^6$Li($\pi,\pi$) and
$^6$Li($\pi,\pi^\prime$) scattering at multiple pion energies.
The data are taken from Ref.~\protect\cite{rit91}.}
\label{fig:xsect6}
\end{figure}

Our results for $^6$Li are displayed in Fig.~\ref{fig:xsect6}.
We see good agreement with the differential cross sections for
elastic scattering.
However, the discrepancies in reproducing the diffractive 
minima at 180 MeV indicate some deficiencies of the simple optical 
potential we have employed.
We also see general agreement with the inelastic scattering to the $(3^+;T=0)$ 
excited state, although some noticeable discrepancies are seen, particularly 
at 100 MeV. 
Nevertheless, it is fair to say that our results agree with the differential
cross sections to a very large extent in both absolute magnitude and 
energy-dependence.
The agreement seen in Fig.~\ref{fig:xsect6} is consistent with the 
$B(E2)$ values listed in Table~\ref{tab:be2}.
The overall agreement is not surprising in view of the ability of the QMC
wave functions to reproduce the elastic and transition form factors in
electron scattering experiments~\cite{WS98}.  
To further improve the agreement with the data and to account for the 
spin observables, it would be necessary to improve the reaction model.
For example, we may have to consider coupled-channel effects and refine 
the optical potential, but this is beyond the scope of the present 
investigation.

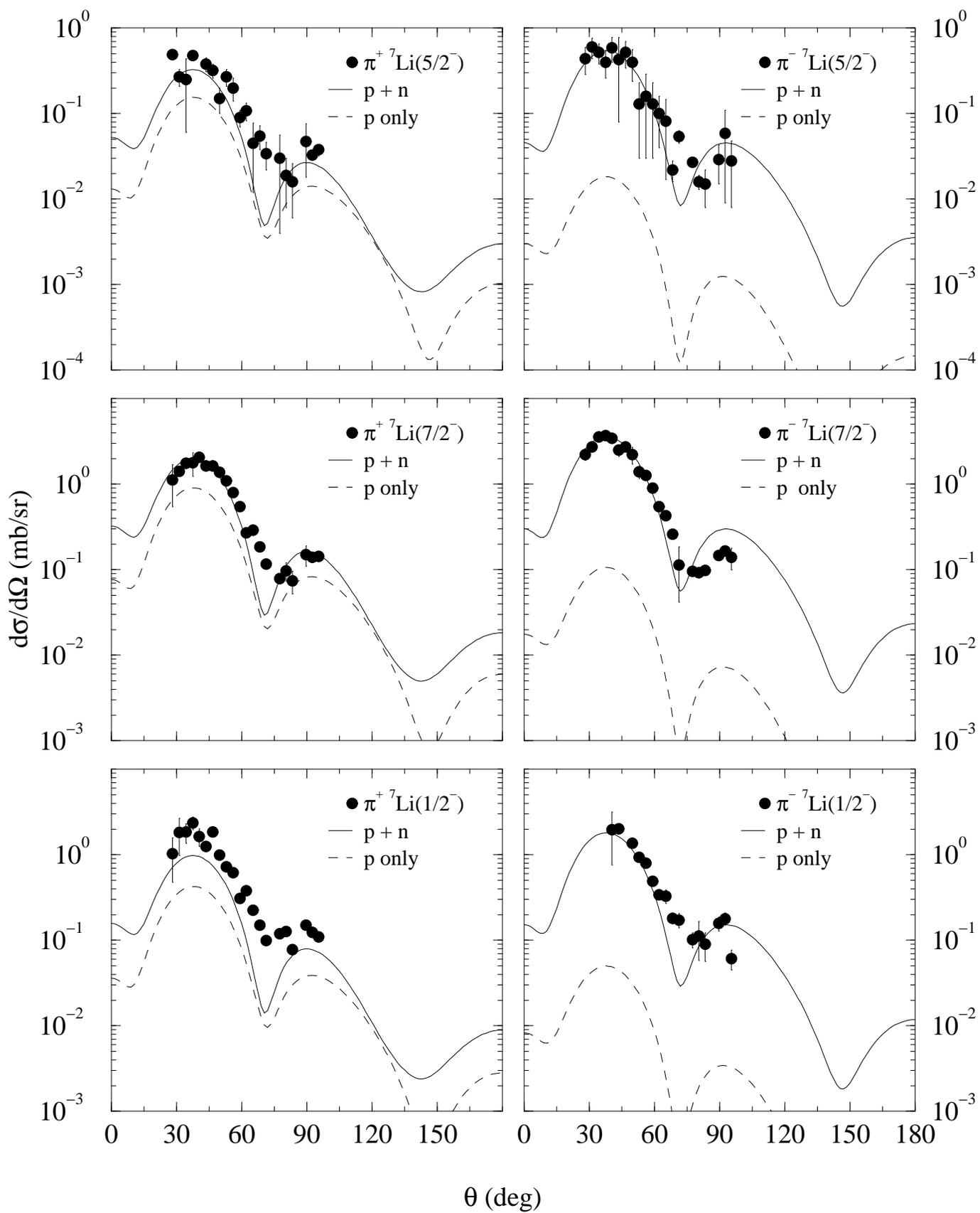
\begin{figure}
\caption{Differential cross sections for $^7$Li($\pi,\pi^\prime$) 
scattering at $E_{\pi}$ = 164 MeV. 
The data are taken from Ref.~\protect\cite{bolger}.}
\label{fig:xsection}
\end{figure}

We next investigate the very old data for 164 MeV $\pi^+$ and $\pi^-$ inelastic 
scattering from a $^7$Li target~\cite{bolger}.
The final states we consider are 
$(J^\pi;T)=(\case{1}{2}^-;\case{1}{2})$ at 0.478 MeV,
$(\case{7}{2}^-;\case{1}{2})$ at 4.63 MeV, and
$(\case{5}{2}^-;\case{1}{2})$ at 6.68 MeV.
Our results are shown in Fig.~\ref{fig:xsection}. 
We see that the predicted cross sections (solid curves) are in excellent 
agreement with the data. 
In the same figure, we also show the contributions obtained using just
the proton excitations.
The agreement with both the $\pi^+$ and $\pi^-$ data is evidently due to the
delicate interplay between the neutron and proton excitations.
If the shell model input given in Ref.\cite{leekurath} is used without 
the enhancement factors $E_n=1.75$ and $E_p=2.5$ for the quadrapole transition
$J(KS)=2(20)$, the predicted cross sections will be a factor of about
5 lower than the data for all of the cases considered here.

In conclusion, we have performed calculations of pion
scattering from $^6$Li and $^7$Li using the nuclear transition
densities predicted by the recent QMC calculations for light
nuclei starting with realistic two-nucleon and three-nucleon potentials.
The predicted cross sections are in very good agreement with the data.
In contrast with the previous calculations using densities from nuclear shell
model, the calculation does not include any enhancement factors.
Because of the strong isospin dependence of the $\pi N$ scattering
t-matrix, the present investigation has probed critically the
predicted neutron transition densities which are not well tested in
electron scattering studies.
Our results suggest that the wave functions predicted by the QMC
calculations are accurate for investigating various nuclear reactions.

It is highly desirable to extend the present work to
re-investigate the very extensive data of pion-nucleus scattering on larger
1p-shell targets and other more complex processes such as pion absorption and 
double-charge-exchange reactions. 
We expect that QMC wave functions for $A=9,10$ nuclei will become available
in the next year.
With the nuclear correlations correctly accounted for by using the wave 
functions predicted by QMC calculations, one now can hope to resolve 
many long-standing problems in intermediate-energy pion-nucleus reactions.

\acknowledgments

We wish to thank D.\ Kurath and S.\ C.\ Pieper for many useful comments.
Our work is supported by the U. S. Department of Energy, Nuclear
Physics Division, under contract No. W-31-109-ENG-38.

\narrowtext

\begin{table}
\caption{$B(E2\uparrow,t_3)$ values in $e^2\cdot$fm$^4$ for different 
transitions in $^6$Li and $^7$Li.  
Experimental values are from Refs.~\protect\cite{yen,vermeer}.  
Experimental uncertainties and Monte Carlo sampling errors are given.}
\begin{tabular}{lddd}
& Experiment & \multicolumn{2}{c}{QMC}  \\
$^A$Z($J_i \rightarrow J_f$) & p & p & n  \\
\tableline
$^6$Li($0^+          \rightarrow 3^+         $) 
& 21.8$\pm$4.8 & 21.1$\pm$0.4 & 21.1$\pm$0.4   \\
\tableline
$^7$Li($\case{3}{2}^-\rightarrow\case{1}{2}^-$) 
&  7.59$\pm$0.10&  5.7$\pm$0.1 & 16.5$\pm$0.3   \\
$^7$Li($\case{3}{2}^-\rightarrow\case{7}{2}^-$) 
& 15.5$\pm$0.8  & 13.2$\pm$0.2 & 34.6$\pm$0.5   \\
$^7$Li($\case{3}{2}^-\rightarrow\case{5}{2}^-$) 
&  4.1$\pm$2.0 &  2.3$\pm$0.1 &  5.4$\pm$0.2   \\
\end{tabular}
\label{tab:be2}
\end{table}

\end{document}